\documentclass[11pt]{article}
\usepackage[hmargin=3cm,vmargin=3cm]{geometry}
\usepackage{llncsdoc}
\usepackage{pdfpages}
\usepackage{color}
\usepackage{fixltx2e} 
\usepackage{mparhack} 
\usepackage{epsfig}
\usepackage[english]{babel}
\usepackage[section]{algorithm}
\usepackage{algpseudocode}
\usepackage{xcolor}
\usepackage{fmtcount} 
\usepackage[cmex10]{amsmath}
\usepackage{amssymb}
\usepackage{amsthm}
\usepackage{mathtools}
\usepackage{amsmath}
\usepackage{wrapfig}
\usepackage{comment}
\usepackage{color}
\usepackage{graphicx}
\usepackage{psfrag}
\usepackage[T1]{fontenc}

\newtheorem{theorem}{Theorem}[section]
\newtheorem{lemma}[theorem]{Lemma}

\newtheorem{corollary}[theorem]{Corollary}
\newtheorem{definition}[theorem]{Definition}
\newtheorem{example}[theorem]{Example}
\newtheorem{remark}[theorem]{Remark}

\newtheorem{fact}[theorem]{Fact}

\newcommand{\BO}{\mathcal{O}}

\begin{document}
	
\title{
Distributed Approximation of Minimum Routing Cost Trees\thanks{This is a full and extended version of \cite{MRCT-SIROCCO}.} 
}

\author{
Alexandra Hochuli
\\
ETH Zurich
\\
hochulia@ethz.ch
\and
Stephan Holzer\footnote{Part of this work was done at ETH Zurich. At MIT the author was supported by the following grants: AFOSR Contract Number FA9550-13-1-0042, NSF Award 0939370-CCF, NSF Award CCF-1217506, NSF Award number CCF-AF-0937274.}
\\
MIT
\\
holzer@mit.edu
\and
Roger Wattenhofer
\\
ETH Zurich
\\
wattenhofer@ethz.ch
}

\maketitle

\begin{abstract}
We study the NP-hard problem of approximating a Minimum Routing Cost Spanning Tree in the message passing model with limited bandwidth (CONGEST model). In this problem one tries to find a spanning tree of a graph $G$ over $n$ nodes that minimizes the sum of distances between all pairs of nodes. In the considered model every node can transmit a different (but short) message to each of its neighbors in each synchronous round. We provide a randomized $(2+\varepsilon)$-approximation with runtime $\BO(D+\frac{\log n}{\varepsilon})$ for unweighted graphs. Here, $D$ is the diameter of $G$. This improves over both, the (expected) approximation factor $\BO(\log n)$ and the runtime $\BO(D\log^2 n)$ stated in~\cite{khan2008efficient}. 

Due to stating our results in a very general way, we also derive an (optimal) runtime of $\BO(D)$ when considering $\BO(\log n)$-approximations as in~\cite{khan2008efficient}. In addition we derive a deterministic $2$-approximation.

\end{abstract}

\section{Introduction}\label{sec:intro}
A major goal in network design is to minimize the cost of communication between any two
vertices in a network while maintaining only a substructure of the network. Despite the fact that a tree is the sparsest substructure of a network it can be surprisingly close to the optimal solution. Every network contains a tree whose total cost of communication between all pairs of nodes is only a factor two worse than the communication cost when all edges in the graph are allowed to be used!

The problem of finding trees that provide a low routing cost is studied since the early days of computing in the 1960s~\cite{scott1969optimal}
and is known to be NP-hard~\cite{johnson1978complexity} on weighted and unweighted graphs\footnote{Even for seemingly simpler versions than those which we study the problem remains NP-hard~\cite{ye2002polynomial}.}. These days networks of computers and electric devices are omnipresent and trees offer easy and fast implementations for applications. In addition, trees serve as the basis for control structures as well as for information gathering/aggregation and information dissemination. This explains why routing trees are computed and used by wide spread protocols such as the IEEE 802.1D standard~\cite{campos2008fast}. When bridging~\cite{wiki2} is used in Local Area Networks (LAN) and Personal Area Networks (PAN), a spanning tree is computed to define the (overlay)
network topology. Finding such a tree with low routing cost is crucial. As~\cite{campos2008fast} demonstrates, current implementations do not perform well under the aspect of optimizing the routing costs and there is the need to find better and faster solutions. The nature of this problem and growth of wired and wireless networks calls for fast and good distributed implementations.

In this paper we present new approaches for distributed approximation of a Minimum Routing Cost Spanning Tree (MRCT) while extending previous work for approximation of those. By doing so we improve both, the round complexity and the approximation factor of the best known (randomized) result in a distributed setting for unweighted graphs. Our main contribution is an algorithm that computes a $\left(2-\frac{2}{n}+\min\left\{\frac{\log n}{D},\alpha(n,D)\right\}\right)$-approximation in time $\BO\left(D+\frac{\log n}{\alpha(n,D)}\right)$ w.h.p.\footnote{A more precise statement can be found in Theorem~\ref{thm:mrct-rand}. This Theorem also considers a generalized version of MRCT.}. Previously, the best known distributed approximation for MRCT~\cite{khan2008efficient} (on weighted graphs) achieved an (expected) approximation-ratio of $\BO(\log n)$ using randomness. The bound on the runtime of the algorithm of~\cite{khan2008efficient} is $\BO(n\log^ 2 n)$ in the worst case -- even when the network is fully connected (a clique). For unweighted graphs, the authors of~\cite{khan2008efficient} specify this runtime to be $\BO(D\log^2 n)$. The distributed algorithms we present in this paper are for unweighted graphs as well\footnote{They extend to graphs with certain realistic weight-functions.} and compared to the (expected) approximation-ratio $\BO(\log n)$ of~\cite{khan2008efficient} we essentially obtain a (guaranteed) approximation-ratio $2+\varepsilon$ in time $\BO(D+\frac{\log n}{\varepsilon})$ w.h.p.. This follows from choosing $\alpha(n,D)=\varepsilon$ for an arbitrary small $\varepsilon>0$. When choosing $\alpha(n,D)=\log n$, we obtain the same approximation ratio as in~\cite{khan2008efficient} in time $\BO(D)$. To be general, we leave the choice of $\alpha(n,D)$ to the reader depending on the application.

Besides this randomized solution we present a deterministic algorithm running in linear time $\BO(n)$ achieving an approximation-ratio of $2$.

\section{Model and Basic Definitions}

Our network is represented by an undirected graph $G=\left(V,E\right)$. Nodes $V$ correspond to processors, computers or routers. Two nodes are connected by an edge from set $E$ if they can communicate directly with each other. We denote the number of nodes of a graph by $n$, and the number of its edges by $m$. Furthermore we assume that each node has a unique ID in the range of $\{1,\dots,2^{\BO\left(\log n\right)}\}$, i.e. each node can be represented by $\BO\left(\log n\right)$ bits. Nodes initially have no knowledge of the graph $G$, other than their immediate neighborhood.

We consider a synchronous communication model, where every node can send $B$ bits of information over all its adjacent edges in one synchronous round of communication. We also consider a modified model, where time is partitioned into synchronized slots, but a message might receive a delay when traversing an edge. This delay might not be uniform but fixed for each edge. In principle it is allowed that in each round a node can send different messages of size $B$ to each of its neighbors and likewise receive different messages from each of its neighbors. Typically we use $B=\BO\left(\log n\right)$ bits, which allows us to send a constant number of node or edge IDs per message.
Since communication cost usually dominates the cost of local computation, local computation is considered to be negligible.
For $B=\BO\left(\log n\right)$ this message passing model is known as CONGEST model~\cite{peleg}.
We are interested in the number of rounds that a distributed algorithm needs to solve some problem. This is the time complexity of the algorithm.

To be more formal, we are interested in evaluating a function $g:\mathbb{G}_n\rightarrow S$, where $\mathbb{G}_n$ is the set of all graphs over $n$ vertices and $S$ is e.g. $\{0,1\}$, $\mathbb{N}$ or $\mathbb{G}_n$, and define distributed round complexity as follows:
\begin{definition}[Distributed round complexity]
Let $\mathcal{A}$ be the set of distributed deterministic algorithms that evaluate a function $g$ on the underlying graph $G$ over $n$ nodes (representing the network). Denote by $R^{dc}\left(A\left(G\right)\right)$ the distributed round complexity (indicated by dc) representing the number of rounds that an algorithm $A\in \mathcal{A}$ needs in order to compute $g\left(G\right)$. We define
	$R^{dc}\left(g\right) = \min_{A \in \mathcal{A}}\max_{G \in \mathbb{G}_n} R^{dc}\left(A\left(G\right)\right)$
to be the smallest amount of rounds/time slots any algorithm needs in order to compute $g$.
\end{definition}
We denote by $R^{dc-rand}_\varepsilon\left(g\right)$ the randomized round complexity of $g$ when the algorithms have access to randomness and compute the desired output with an error probability smaller than $\varepsilon$. By w.h.p. (with high probability) we denote a success probability larger than $1-1/n$.

The unweighted shortest path in $G$ between two nodes $u$ and $v$ is a path with minimum number of edges among all $(u,v)$-paths. Denote by $d_G\left(u,v\right)$ the unweighted distance between two nodes $u$ and $v$ in $G$ which is the length of an unweighted shortest $(u,v)$-path in $G$. We also say $u$ and $v$ are $d_G(u,v)$ hops apart. By $\omega_G:E\rightarrow \mathbb{N}$ we denote a graph's weight function and by $\omega_G(e)$ the weight of an edge in $G$. By 
$\omega_G(u,v):=\min_{\{P|P\text{ is }(u,v)\text{-path in }G\}}\sum_{\text{edges } e \text{ in }P}\omega_G(e)$
 we define the weighted distance between two nodes $u$ and $v$, that is the weight of a shortest weighted path in a graph $G$ connecting $u$ and $v$\footnote{ Note that in the context of MRCT, $\omega$ often corresponds to the cost of an edge. In the literature the routing cost between any node $u$ and $v$ in a given spanning tree $T$ of $G$ is usually denoted by $c_T\left(u, v\right)$, while in generalized versions of MRCT, the weight of an edge can be different from the cost. In this paper we use $\omega_T\left(u,v\right)=c_T\left(u, v\right)$.}. 

The time-bounds of our algorithms as well as those of previous algorithms depend on the diameter of a graph. We also use the eccentricity of a node.
\begin{definition}[Eccentricity, diameter]
The \emph{weighted eccentricity} $ecc_{\omega_G}\left(u\right)$ in $G$ of a node $u$ is the largest weighted distance to any other node in the $G$, i.e. $ecc_{\omega_G}\left(u\right):=\max_{v\in V} \omega_G\left(u,v\right)$. The \emph{weighted diameter} $D_{\omega}\left(G\right):=\max_{u\in V} ecc_{\omega_G}(u):=\max_{u,v\in V} \omega_G\left(u,v\right)$ of a graph $G$ is the maximum weighted distance between any two nodes of the graph. The \emph{unweighted diameter} (or hop diameter) $D_{h}\left(G\right):=\max_{u,v\in V} \min_{\{P|P\text{ is }(u,v)\text{-path}\}}|P|$ of a graph $G$ is the maximum number of hops between any two nodes of the graph. Here $|P|$ indicates the number of edges on path $P$.
\end{definition}
We often write $D_{\omega}$ and $D_h$ instead of $D_{\omega}(G)$ and $D_h(G)$ when we refer to the diameter of a graph $G$ in context. Observe that $D_h=D_\omega$ for unweighted graphs. 

Finally, we define the problems that we study.

\begin{definition}[$S$-Minimum Routing Cost Tree ($S$-MRCT)]
Let $S$ be a subset of the vertices $V$ in $G$. The $S$-routing cost of a subgraph $H$ is defined as $RC_S\left(H\right):=\sum_{u, v\in S} \omega_H\left(u, v\right)$ and denotes the routing cost of $H$ with respect to $S$. An $S$-MRCT is a subgraph $T$ of $G$ that is a tree, contains all nodes $S$ and has minimum $S$-routing cost $RC_S\left(T\right)$ among all spanning trees of $T$. 
\end{definition}
This is a generalization of the MRCT problem~\cite{WuLBCRT99}. According to this definition $V$-MRCT (i.e. $S=V$) and MRCT of~\cite{WuLBCRT99} are equivalent. Therefore all results are valid for the classical MRCT problem when choosing $S:=V$.

In this paper we consider approximation algorithms for these problems. Given an optimization problem $P$, denote by $OPT$ the cost of the optimal solution for $P$ and by $SOL_A$ the cost of the solution of an algorithm $A$ for $P$. We say $A$ is $\rho$-approximative for $P$ if $OPT\le SOL_A \le \rho\cdot OPT$ for any input.

\begin{fact}\label{fact:ecc-approx-diam}
The eccentricity of any node is a good approximation of the diameter. For any node $u\in V$ we know that $ecc_{\omega_G}\left(u\right)\leq D_{\omega}\left(G\right) \leq 2\cdot ecc_{\omega_G}\left(u\right)$.
\end{fact}

\section{Our Results}\label{sec:results}
In Section~\ref{sec:final} we prove the following two theorems.
\begin{theorem}\label{thm:mrct}
In the CONGEST model, the deterministic algorithm proposed in Section~\ref{sec:final} needs time $\BO\left(|S|+D_\omega\right)$ to compute a $\left(2-2/|S|\right)$-approximation for $S$-MRCT when using either uniform weights for all edges or a weight function $\omega(e)$ that reflects the delay/edge traversal time of edge $e$.
\end{theorem}

\begin{theorem}\label{thm:mrct-rand}
Let $\alpha(n,D_\omega)$ be some function in $n$ and $D_\omega$. The randomized algorithm proposed in Section~\ref{sec:final} computes w.h.p. a 
$\left(2-\frac{2}{|S|}+\min\left\{\frac{\log n}{D_\omega},\alpha(n,D_\omega)\right\}\right)$-approximation for $S$-MRCT in the CONGEST model in time $\BO\left(D_\omega+\frac{\log n}{\alpha(n,D_\omega)}\right)$ when using either uniform weights for all edges or a weight function $\omega(e)$ that reflects the delay/edge traversal time of edge $e$.
\end{theorem}

We emphasize that the analysis of~\cite{wong1980worst} yields a $2$-approximation when compared to the routing cost in the original graph\footnote{Note that most other approximation algorithms are with respect to  the routing cost of a minimal routing cost tree of the graph. In Section~\ref{sec:cant} we provide an example that shows that sometimes even no subgraph with $o(n^2)$ edges exists that yields better approximations to the routing cost in the original graph than the trees presented here. From this we conclude that algorithms that compare their result only to the routing cost of the minimum routing cost tree do not always yield better results than those presented here.} and that we modify this analysis.

\section{Related Work}\label{sec:related}
Minimum Routing Cost Trees are also known as uniform Minimum Communication Cost Spanning Trees~\cite{peleg2002low,reshef1999approximating} and shortest Total Path Length Spanning Trees~\cite{wu2000approximation}. Furthermore the MRCT problem is a special case of the Optimal Network Problem, first studied in the 1960s by~\cite{scott1969optimal} and later by~\cite{dionne1979exact}. In~\cite{wong1980worst} Wong presented heuristics and approximations to the Optimal Network Problem with a restriction that makes the problem similar to the MRCT problem and obtained a $2$-approximation. In~\cite{johnson1978complexity} it is shown that this restricted version, which Wong studied on unweighted graphs, is NP-hard as well. It seems that earlier the authors of~\cite{hu1974optimum} formulated a similar problem under the name ''Optimum communication spanning tree" where in addition to costs on edges, we are given a requirement-value $r_{u,v}$ for each pair of vertices that needs to be taken into account when computing the routing cost. In this setting one wants to find a tree $T$ such that $\sum_{u,v\in V} r_{u,v}d_T(u,v)$ is minimized. In~\cite{WuLBCRT99} it is argued that for metric graphs, the results by~\cite{bartal1996probabilistic,bartal1998approximating,charikar1998rounding} yield a $\BO(\log n \log\log n)$-approximation to this problem. Using a result presented in~\cite{fakcharoenphol2003tight}, this can be improved to be an $\BO(\log n)$-approximation. In~\cite{khan2008efficient} it is shown how to implement this result in a distributed setting. They state their result depending on the shortest path diameter $D_{sp}(G):=\max_{u,v\in V}\{|P| \; | P$ is a shortest weighted $(u,v)\text{-path}\}$  of a graph. This diameter represents the maximum number of hops of any shortest weighted path between any two nodes of the graph. The authors of~\cite{khan2008efficient} obtain a randomized approximation of the MRCT with expected approximation-ratio $\BO(\log n)$ in time $\BO\left(D_{sp}\cdot \log^2 \left(n\right)\right)$. Observe that this might be only a $\BO(n\log^2 n)$-approximation even in a graph with $D_h=1$ and $D_{sp}=n-1$, such as a clique where all edges have weight $n$ except $n-1$ edges of weight $1$ forming a line as a subgraph.\footnote{According to~\cite{WuLBCRT99} it is NP-hard to find an MRCT in a clique.} In our distributed setting we know that it is hard to approximate an MRCT due to Theorem~\ref{thm:randomized_approximation}. 
\begin{theorem}[Version of Theorem 5.1. of~\cite{sarma2012distributed}]\label{thm:randomized_approximation}
For any polynomial function $\alpha\left(n\right)$, numbers $p$, $B\geq 1$, and $n\in \{2^{2p+1}pB, 3^{2p+1}pB, \ldots\}$, there exists a constant $\varepsilon>0$ such that in the CONGEST model any distributed $\alpha(n)$-approximation algorithm for the MRCT problem whose error probability is smaller than $\varepsilon$ requires $\Omega\left(\left(\frac{n}{pB}\right)^{\frac{1}{2}-\frac{1}{2\left(2p+1\right)}}\right)$ time on some $\Theta\left(n\right)$-vertex graph of diameter $2p+2$.
\end{theorem}

For certain realistic weight-functions our randomized algorithm breaks this $\Omega(\sqrt{n}+D)$-time lower bound. This is no contradiction, as the construction of~\cite{sarma2012distributed} heavily relies on being able to choose highly different weights, which might not always appear in practice: in current LAN/PAN networks, weights (delays) usually differ only by a small factor. In case the weights are indeed the delay-times, the runtime of our algorithm just depends on the maximal delay that occurs between any two nodes in the network. Observe that also the runtime of the algorithm of~\cite{khan2008efficient} stated for arbitrary weight functions does not contradict this approximation lower bound. The algorithm's runtime depends on the shortest path diameter $D_{sp}$, which is $\Theta(\sqrt{n}+D)$ in the worst case graphs provided in~\cite{sarma2012distributed}. Finally we want to point out that for weighted graphs it might be possible to combine the recent result of~\cite{nanongkai2014distributed} with the techniques developed in this paper. This might improve over the approximation factor of~\cite{khan2008efficient} for weighted graphs while getting a better runtime in some cases.

Related work in the non-distributed setting includes~\cite{WuLBCRT99}, where a PTAS to find the MRCT of a weighted undirected graph is presented. It is shown how to compute a $\left(1+2/(k+1)\right)$-approximation for any $k\geq 1$ in time $\BO\left(n^{2k}\right)$.

Already for $k\geq 2$, the PTAS of~\cite{WuLBCRT99} yields a time bound of $\BO(n^4)$ and we cannot expect to obtain a distributed algorithm running in time $o(n^2)$ since we can only hope for a distributed speedup by at most $n+m$. Setting $k=1$ yields a $2$-approximation in (sequential) time $\BO(n^2)$ and we could hope at most to obtain a distributed runtime of $\BO(n^2/(n+m))$ from this. While one could try to transform this algorithm into our distributed setting, our algorithm based on~\cite{wong1980worst} is simpler and yields the same approximation ratio of $2$. In addition we derive a fast randomized version from this.

Further related work on parallel approximations for MRCT in RNC circuits was published in~\cite{chang2007parallelized}. Here, RNC abbreviates the complexity class \emph{Randomized Nick's Class}, which consists of all decision problems decidable by uniform Boolean circuits with a polynomial number of gates of at most two inputs and depth $\BO(\log n)$.  Wu considered in~\cite{ye2002polynomial} the version of MRCT, where one is only interested in minimizing the routing cost from two source vertices to all nodes in the network and is hence denoted by $2$-MRCT. He does this in a non-distributed setting and proves NP-hardness. He also presents a polynomial time approximation scheme (PTAS) for this version of the problem. Note that the $2$-MRCT problem is different from the special case of the $S$-MRCT problem with $|S|=2$, where only a tree that connects nodes in $S$ should be found. 
Recent speedups on exact solutions and heuristics for the MRCT problem can be found in~\cite{campos2008fast,fischetti2002exact}.

There is also a large body of work on Low Stretch Spanning Trees~\cite{DBLP:conf/focs/AbrahamBN08,elkin2008low,peleg2002low}. The stretch for an edge $(u,v)$ in $E$ using spanning tree $T$ of $G$ is defined to be $stretch_T(u,v):=\omega_T(u,v)/\omega_G(u,v)$ and the average stretch is $avestr(G,T):=\frac{1}{|E|}\sum_{(u,v)\in E}stretch_T(u,v)$. A tree with maximum stretch $\alpha$ yields an $\alpha$-approximation to the routing cost in $G$. However, the maximal stretch can be high and thus in general does not yield better bounds on the routing cost than the trees presented here. Still, algorithms that yield good bounds on the average stretch are known -- $\BO(\log^2n \log\log n)$ can be achieved and was lower bounded by $\Omega(\log n)$ in~\cite{elkin2008low}. Average stretch and routing cost quality are unrelated.

\section{Trees that $2$-Approximate the Routing Cost}\label{sec:det-ana}
The main structure we need in this section are shortest path trees:
\begin{definition}[Shortest path tree]
A shortest path tree (SP-tree) rooted in a node $v$, is a tree that connects any node $u$ to the root $v$ by a shortest path in $G$. In unweighted graphs, this is simply a breadth first-search tree.
\end{definition}
Previously it was known due to Wong~\cite{wong1980worst}, Theorem $3$, that there is an SP-tree, which $2$-approximates the routing cost of an MRCT. We restate this result by using an insight stated in Wong's analysis such that this tree not only $2$-approximates the routing cost $RC_V(T)$ of an MRCT $T$ of $G$ (which is a $V$-MRCT) as Wong stated it, but even yields a $2$-approximation of the routing cost $RC_V(G)$ when using shortest paths in the network $G$ itself. Thus, on average the distances between two pairs in the tree are only a factor $2$ worse than the distances in $G$. 

The algorithm that corresponds to Wong's analysis computes and evaluates $n$ SP-trees, one for each node in $V$. We show, that for the $S$-MRCT problem it is sufficient to consider only those shortest path trees rooted in nodes of $S$. At the same time, a slightly more careful analysis yields a slightly improved approximation factor of $2-2/|S|$, which is of interest for small sets $S$. Before we start, we define a useful measure for the analysis.
\begin{definition}[Single source routing cost]
By $SSRC_S\left(v\right):=\sum_{u\in S}\omega_G\left(v,u\right)$ we denote the sum of the single source routing costs from node $v$ to every other node in $S$ by using edges in $G$.
\end{definition}
Note that for simplicity we defined an SP-tree to contain all nodes of $V$. However, one could also consider the subtree where all leaves are nodes in $S$. The measures $RC_S$ and $SSRC_S$ would not change, as any additional edges are never used by any shortest paths and thus do not contribute to the $S$-routing cost of the tree. Such a tree can easily be obtained from the tree we compute.
\begin{theorem}\label{thm:det-ana}
Let $|S|$ be at least $2$. In weighted graphs, the SP-tree $T_v$ rooted in a node $v$ with minimal single source routing cost $SSRC_S(v) = \min_{u \in S} SSRC_S(u)$ over all SP-trees rooted in nodes of $S$ is a $\left(2-2/|S|\right)$-approximation to the $S$-routing cost $RC_S(G)$ in $G$.
\end{theorem}
\begin{corollary}\label{cor:improved}
In weighted graphs, an SP-tree with minimum routing cost over all SP-trees rooted in nodes of $S$ is a $\left(2-2/|S|\right)$-approximation to an $S$-MRCT.
\end{corollary}
The proof of this theorem uses and modifies the ideas of the proof of Theorem 3 in~\cite{wong1980worst}. The following proof is an adapted version of this proof. 

\begin{proof}
Let $v$ be the node for which the SP-tree $T_{v}$ has minimal single source routing cost with respect to $S$ among all SP-trees, that is $v:=arg\min_{v\in V}SSRC_S\left(v\right)$. 

The cost of connecting a node $u\neq v$ to all other nodes in $S$ using edges in $T_{v}$ is upper bounded by $\left(|S|-2\right)\cdot \omega_G\left(v,u\right)+SSRC_S\left(v\right)$. This essentially describes the cost of connecting $u$ to each other node by a path via the root $v$ and using edges in $T_v$. Therefore the total routing cost $RC_S\left(T_{v}\right)$ for $S$ using the network $T_v$ can be bounded by
\begin{align*}
&RC_S\left(T_{v}\right) \leq SSRC_S\left(v\right) + \sum_{v\neq u\in S}\left(\left(|S|-2\right)\cdot \omega_G\left(v,u\right)+SSRC_S\left(v\right)\right).
\end{align*}
As $|S|\geq 2$, this can be further transformed and bounded to be 
\begin{eqnarray*}
& = & |S| \cdot SSRC_S\left(v\right) + \left(|S|-2\right)\sum_{u\in S}\omega_G\left(v,u\right)
\\
&=&
|S|\cdot SSRC_S\left(v\right) + \left(|S|-2\right)\cdot SSRC_S\left(v\right)
\\
&=& (2-2/|S|)\cdot|S|\cdot SSRC_S\left(v\right)
\\
&\leq&
(2-2/|S|)\cdot\sum_{u\in S}SSRC_S\left(u\right).
\end{eqnarray*}
Where the last bound follows, as $SSRC_S\left(v\right)$ is minimal among all $SSRC(u)$ for $u\in S$. Since $\sum_{u\in V}SSRC_S\left(u\right)$ is the same as $RC_S\left(G\right)$, we obtain that $RC_S\left(T_{v}\right)\leq 2 RC_S\left(G\right)$.
\end{proof}

\section{Considering few Randomly Chosen SP-Trees is Almost as Good}\label{sec:rand-ana}
We show that when investigating a small subset of all SP-trees chosen uniformly at random, with high probability one of these trees is a good approximation as well.
\begin{lemma}\label{lem:rand-ana}
Let $\beta(n,D)$ be a positive function in $n$ and $D$ and define $\gamma:=\left\lceil\frac{2-2/|S|}{\beta(n,D)}\right\rceil+1$. Assume $S\subseteq V$ is of size at least $\gamma\ln n$. Let $S'$ in turn be a subset of $S$ chosen uniformly at random among all subsets of $S$ of size $\gamma\ln n$. Let $v\in S'$ be a node such that $SSRC_S(v)=\min_{u\in S'}SSRC_S(u)$. Then $RC_S(T_v)\leq (2-2/|S|+\beta(n,D))RC_S(G)$.   
\end{lemma}
\begin{proof}
For simplicity, without loss of generality we assume that $|S|$ is a multiple of $\gamma$. Denote by $v_1,\dots,v_{|S|}$ the nodes in $S$ such that $SSRC_S(v_1)\leq SSRC_S(v_2)\leq \cdots \leq SSRC_S(v_{|S|}).$ That is they are ordered corresponding to their single source routing costs. We say a node $v$ is good, if the corresponding SP-tree $T_v$ is among the $1/\gamma$-fraction of the SP-trees with lowest single source routing cost\footnote{Due to the choice of $\gamma:=\left\lceil\frac{2-2/|S|}{\beta(n,D)}\right\rceil+1$ a good tree is among the $n\beta(n,D)$ cheapest trees.} . Therefore $v$ is good if $SSRC_S(v)\leq SSRC_S(v_{|S|/\gamma})$ with respect to the above order of the trees.

First we prove that w.h.p. set $S'$ contains a good node. Second we prove, that the corresponding SP-tree yields the desired approximation ratio.

\emph{1) Probability analysis:} We know that $Pr_{v\in S}[v$ is good$]=1/\gamma$. Furthermore each node $v\in S$ is included in set $S'$ independent of the other nodes. Therefore we can conclude that the probability that at least one of the nodes $v$ in $S'$ is good is $1-\left(1-\frac{1}{\gamma}\right)^{|S'|}=1-\left(1-\frac{1}{\gamma}\right)^{\gamma\ln n} > 1- 1/n$ and thus high.

\emph{2) Approximation-ratio analysis:} Let $v_i$ be a good node. 
As in the proof of Theorem~\ref{thm:det-ana} we know that 
$
RC_S(T_{v_i})\leq (2-2/|S|)\cdot |S|\cdot SSRC_S(v_i).\label{eqn:1}
$.
 As $RC_S(G)=\sum_{u\in S} SSRC_S(u)$ and $v_i$ is good, we can conclude that 
$SSRC_S(v_i)\leq \frac{1}{(1-1/\gamma)\cdot|S|}\cdot RC_S(G)$
as there are at most $(1-1/\gamma)|S|$ nodes $v_j$ with $SSRC_S(v_j) \geq SSRC_S(v_i)$.
Equality is approached in the worst case, where $j:=|S|/\gamma$ and $SSRC_S(v_j)=0$ for each $j<i$ and $SSRC_S(v_i)=SSRC_S(v_j)$ for all $j\geq i$. 

Combined with Bound (\ref{eqn:1}) it follows that 
$RC_S(T_{v_i})\leq \frac{2-2/|S|}{1-1/\gamma}\cdot RC_S(G).$ Due to the choice of $\gamma$ we conclude the statement of the Lemma.

\end{proof}

\section{How to Compute the Routing Cost of many SP-trees in Parallel}

In Theorem~\ref{thm:det-ana} (and Lemma~\ref{lem:rand-ana}) we demonstrated that an SP-tree $T_v$ with minimum single source routing cost yields a $2$-approximation for $RC_S(G)$. The single source routing cost of a tree can be computed by computing distances between the root of a tree and nodes in $S$. However, instead of finding an SP-tree with smallest single source routing cost the literature usually considers finding an SP-tree with smallest routing cost. This is done e.g. in~\cite{wong1980worst}. The reason for this is that the bound in the proof of Lemma~\ref{thm:det-ana} is not sharp when using the single source routing cost. To see this, we recall that while obtaining the bound, one approximates the distance between two nodes in the tree by adding up their distance to the root. Thus the bound considers the single source routing cost of an SP-tree. Compared to this, the routing cost takes the actual distance of the two nodes in an SP-tree into account. An explicit example for a graph that contains a node $u$ such that $RC_S(T_u)<RC_S(T_v)$, where $T_v$ has minimum single source routing cost is given in Example~\ref{ex:example22}. Like in~\cite{wong1980worst} we focus on this more powerful version of finding a tree of small routing cost. 

\begin{example}\label{ex:example22}
Consider the graph $G$ in Figure~\ref{fig:example22}. The tree $T_u$ has smallest single source routing cost $SSRC_V(T_u)=5$. At the same time the tree $T_v$ has single source routing cost $SSRC_V(T_v)=7$, while its routing cost $RC_V(T_v)=32$ is lower than the routing cost $RC_V(T_u)=36$ of $T_u$. Note that the actual routing cost in $G$ is $RC_V(G)=27$. 
\begin{figure}[!htb]
\begin{center}
\includegraphics[scale=0.5]{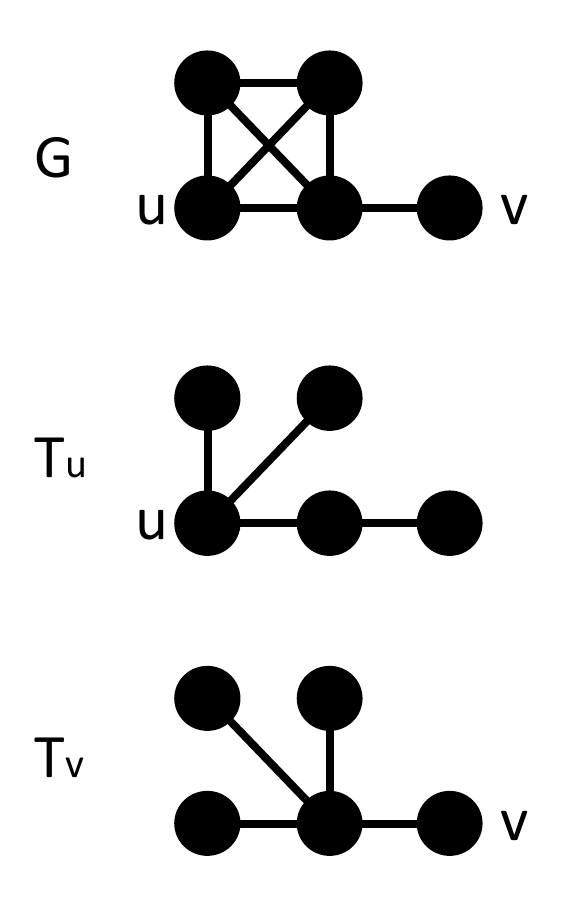}
\caption{Unweighted graph $G$ with distinguished vertices $u$ and $v$ as well as SP-trees $T_u$ and $T_v$ corresponding to Example \ref{ex:example22}.}
\label{fig:example22}
\end{center}
\end{figure} 
\end{example}

\begin{lemma}\label{lem:RC-comp}
Let $S:=\{v_1,\dots,v_{|S|}\}$ be a subset\footnote{Note that $S$ used here can be e.g. $S$ as in Section~\ref{sec:det-ana} or the smaller set $S'$ as in Section~\ref{sec:rand-ana}.} $S\subseteq V$ of all nodes of a graph. Then we can compute the values $RC_S(T_{v_1}),\dots,RC_S(T_{v_{|S|}})$ in time $\BO(D_\omega+|S|)$ when using either uniform weights for all edges or a weight function implied by the delay/edge traversal time.
\end{lemma}

The proof of this lemma can be found at the end of this section. First, we describe our algorithm that is used to prove this lemma. In Part $1$ of this algorithm we start by computing SP-trees $T_v$ for each $v\in S$. A pseudocode for this algorithm can be found as Algorithm~\ref{alg:S-MRCT-Part-1}. Part 2 deals with computing the routing cost of a single tree and is described later in this section.

We start by noting that for the weight functions we consider an SP-tree is just a Breath First Search tree (BFS-tree). This part is essentially the same as in the $S$-SP algorithm of~\cite{holzer2012optimal} extended to edge-weights derived from the delays to send a message. We also store some additional data that is used later in Algorithm~\ref{ALGRAND} to compute routing costs but was not needed for the $S$-SP computation in~\cite{holzer2012optimal}. In Algorithm~\ref{ALGRAND}, for each node $v\in S$ an SP-tree $T_v$ is constructed using what we call delayed breadth first search (DBFS). By DBFS we think of a breadth first search, where traversing edge $(u,u')$ takes $\omega_G(u,u')$ time slots. In the end each node $u$ in the graph knows $\omega_G\left(u,v\right)$. In addition each node $u$ knows for each $v\in S$ its parent in the corresponding tree $T_v$. Furthermore node $u$ knows at what time the DBFS, that computed $T_v$, sent its message to $u$ via $u$'s parent. During Algorithm~\ref{ALGRAND}, these timestamps are used to compute the routing cost of all these trees in time $\BO\left(|S|+D_\omega\right)$.

\begin{algorithm}[!h]
\begin{algorithmic}[1]
\State $L:=\emptyset$; $\omega_u:=\{0,0,\dots,0\}$; $L_{delay}:=\emptyset$;
\State $\tau:=\{\infty,\infty,\dots,\infty\}$  // **new**
\If{$u\in S$}
\State $L:=\{u\}$;
\State $\omega_u\left(u\right):=0$;
\State $\tau\left(u\right):=0$; // **new**
\EndIf
\State $L_1,\dots,L_{\delta\left(u\right)}:=L$;
\If{ $u$ equals $1$}
\State \textbf{compute} $D_\omega':=ecc(u)$; //** According to Fact~\ref{fact:ecc-approx-diam}, $D_\omega$ is smaller than $2\cdot D_\omega'$.
\State \textbf{broadcast} $D_\omega'$;
\Else
\State \textbf{wait until} $D_\omega'$ was \textbf{received};
\EndIf
\algstore{myalg}
\end{algorithmic}\caption{Computing $SSRC_S(v)$ for each $v\in S$ Part 1 (executed by node $u$)}\label{alg:S-MRCT-Part-1}
\end{algorithm}
\newpage
\begin{algorithm} [!h]                    
\begin{algorithmic} [1]                  
\algrestore{myalg}
\State //** Compute $S$ shortest path trees
\For{$t=1,\dots,|S|+2\cdot D_\omega'$}
\For{$i=1,\dots,\delta\left(u\right)$}
\State $(l_i,\omega_i):=\left\{\renewcommand{\arraystretch}{2}
	\begin{array}{l l}
		\ \ \bot & \quad \text{: if $L_i\setminus \cap L_{delay}=\emptyset$} \\
		\renewcommand{\arraystretch}{1}
		\begin{tabular}{l}
		    $\arg\min\left\{v\in L_i\setminus L_{delay}|\right.$\\
				$\left.\tau[v]+\omega_G(u,v)\geq t\right\}$
		\end{tabular}
		& \quad \text{: else}
	\end{array}\right.$
\EndFor
\State within one time slot:
\newline \hspace*{0.8cm}\textbf{if} $l_1\neq \bot$ \textbf{then send} $\left(l_1,\omega_u[l_1]+\omega_G\left(u,u_1\right)\right)$ to neighbor $u_1$;
\newline  \hspace*{0.8cm}\textbf{receive} $\left(r_1,\omega_1\right)$ from $u_1$;
\newline  \hspace*{0.8cm}\textbf{if} $l_2\neq \bot$ \textbf{then send} $\left(l_2,\omega_u[l_2]+\omega_G\left(u,u_2\right)\right)$ to neighbor $u_2$;
\newline  \hspace*{0.8cm}\textbf{receive} $\left(r_2,\omega_2\right)$ from $u_2$;
\newline  \hspace*{0.8cm} $\vdots$
\newline  \hspace*{0.8cm}\textbf{if} $l_{\delta\left(u\right)}\neq \bot$ \textbf{then send} $\left(l_{\delta\left(u\right)},\omega_u[l_{\delta\left(u\right)}]+\omega_G\left(u,l_{\delta\left(u\right)}\right)\right)$ to neighbor $u_{\delta\left(u\right)}$; \newline  \hspace*{0.8cm}\textbf{receive} $\left(r_{\delta\left(u\right)},\omega_{\delta\left(u\right)}\right)$ from $u_{\delta\left(u\right)}$;
\State $R:=\{r_i|r_i<l_i$ and $i\in 1\dots \delta(u)\}\setminus L$
\State $s:=\left\{
  \begin{array}{l l}
    \infty & \quad \text{if $L_{delay} = \emptyset$}\\
    \min(L_{delay}) & \quad \text{else}
  \end{array} \right.$
\If{$s\leq \min(R) \textbf{ and } s<\infty$}
	\State $L_{delay}:=L_{delay}\setminus\{s\}$;
\EndIf
\For{$i=1,\dots,\delta\left(u\right)$}
\If{$r_i<l_i$}
\State //** $T_{l_i}$'s message is delayed due to $T_{r_i}$.
\If{$r_i\notin L$}
\State $\tau[r_i]:=t$; // **new**
\State $\omega_u[r_i]=\omega_i$;
\State $L:=L\cup\{r_i\}, L_1:=L_1\cup\{r_i\}, L_2:=L_2\cup\{r_i\},  
\newline \hspace*{1.9cm}\dots L_{i-1}:=L_{i-1}\cup\{r_i\}, L_{i+1}:=L_{i+1}\cup\{r_i\}, \dots L_{\delta\left(u\right)}:=L_{\delta\left(u\right)}\cup\{r_i\}$;
\If{$\min(R)<r_i \textbf{ or } s<r_i$}
\State $L_{delay}=L_{delay}\cup\{r_i\}$
\EndIf
\State $parent\_in\_T_{r_i}:=$ neighbor $i$;
\EndIf
\Else
\State $L_i:=L_i\setminus \{l_i\}$; //**  $T_{l_i}$'s message was successfully sent to neighbor $i$.
\EndIf
\EndFor
\EndFor
\end{algorithmic}
\end{algorithm}
\newpage

\begin{remark}
Compared to Algorithm $S$-SP presented in~\cite{holzer2012optimal} we added Lines 2, 6 and 26 in Algorithm~\ref{alg:S-MRCT-Part-1} and extended the algorithm to certain delay functions as mentioned above (the proof in~\cite{holzer2012optimal} can be naturally extended to those.) By doing so, we can store in $\tau[v]$ the time when a message of the computation of tree $T_v$ was received the first time (via edge $parent\_in\_T_v$). In the end, $\omega_u[v]$ stores the distance $\omega_G\left(v,u\right)$ to $v$ and $parent\_in\_T_{v}$ indicates the first edge of a $(u,v)$-path witnessing this. 
\end{remark}

Despite its similarity to algorithm $S$-SP in~\cite{holzer2012optimal}, we describe Algorithm~\ref{alg:S-MRCT-Part-1} in more detail for completeness. For the simplicity of the writeup, we refer to $u$ not only as a node, we use $u$ to refer to $u$'s ID as well. Each node $u$ stores $\delta\left(u\right)$ sets $L_i$, one for each of the $\delta\left(u\right)$ neighbors $u_1,\dots,u_{\delta\left(u\right)}$ of $u$, and the sets $L$  and $L_{delay}$ to keep track of which messages were received, transmitted or need to be delayed. At the beginning, if $u\in S$, all these sets contain just $u$, else they are empty (Lines 1--7). Set $L_{delay}$ is always initialized to be empty. Furthermore $u$ maintains an array $\omega_u$ that eventually stores at position $v$ (indicated by $\omega_u[v]$) the distance $\omega_G\left(u,v\right)$ to node $v$. Initially $\omega_u[v]$ is set to infinity for all $v$ and is updated as soon as the distance is known (Line 27). In each node $u$, array $\tau$ stores at position $v$ the time when a message of the computation of tree $T_v$ was received the first time in $u$. At any time, set $L$ contains all node IDs corresponding to the tree computations (where each node with a stored ID is the root initiating the computation of such a tree) that already reached $u$ until now.
The set $L_{delay}$ contains all root IDs that reached $v$ until time $t$ but are marked to be delayed before forwarded. This ensures that we indeed compute BFS-trees.

Set $L_i$ contains all IDs of $L$ except those that could be forwarded successfully to neighbor $u_i$ in the past. We say an ID $l_i$ is forwarded successfully to neighbor $u_i$, if $u_i$ is not sending a smaller ID $r_i$ to $u$ at the same time.

To compute the trees in Algorithm~\ref{alg:S-MRCT-Part-1}, the unique node with ID $1$ computes $D_\omega'$ and thus a $2$-approximation to the distance-diameter $D_\omega$. This value is subsequently broadcast to the network (Lines 8--12). Then the computation of the $|S|$ trees starts and runs for $|S|+2D_\omega'$ time steps. Lines 14--17 make sure that at any time the smallest ID, that is not marked to be delayed and was not already  forwarded successfully to neighbor $u_i$ is sent to $u_i$ together with the length of the shortest $(v,u_i)$-path that contains $u$. In Line 18 we define the set $R$ of all IDs that are received successfully in this time slot for the first time. This set is then used to decide whether to remove an ID $s$ from $L_{delay}$ in Lines 20 and 21, since all IDs that cause a delay to $s$ are transmitted successfully by now. ID $s$ is computed in Line 20. ID $s$ is the smallest element of $L_{delay}$ and is removed from $L_{delay}$ if no other ID smaller than $s$ was received successfully for the first time in this timeslot.

If a node ID $r_i$ was received successful for the first time (verified in Lines 23 and 25), we update $\tau[r_i]$ and $\omega_u[r_i]$, add $r_i$ to the according lists (Lines 28--30) and remember who $u$'s parent is in $T_{r_i}$ (Line 31). In case the ID $v$ was received the first time from several neighbors, the algorithm as we stated it chooses the edge with lowest index $i$. On the other hand if we did not successfully receive a message from neighbor $u_i$ but sent successfully a message to neighbor $u_i$, the transmitted ID is removed from $L_i$ (Line 33).
\begin{lemma}\label{lem:S-MRCT-Part-1-Runtime}
Algorithm~\ref{alg:S-MRCT-Part-1} computes an SP-tree $T_v$ for each $v\in S$ in time $\BO(|S|+D_\omega)$.
\end{lemma}
\begin{proof}
This is essentially Theorem 6.1. in~\cite{holzer2012optimal-merge} stated for Algorithm~\ref{alg:S-MRCT-Part-1} instead of Algorithm $S$-SP of~\cite{holzer2012optimal-merge}. Those parts of the two algorithms which contribute to the runtime and correctness are equivalent. 
\end{proof}

Now Part 2 of our algorithm calculates the routing cost of each tree $T_v$ in parallel in time $\BO(D_\omega+|S|)$.
A pseudocode of this algorithm is stated in  Algorithm~\ref{ALGRAND}.

To compute the routing cost of a tree, we look at each edge $e$ in each tree $T_v$ and compute the number of $(v,w)$-paths in $T_v$ that contain the edge $e$, for $v,w\in S$. The sum of these numbers for each edge in a tree is the tree's routing cost. Given a tree $T$, for each edge $e$ in $T$, the edge partitions the tree into two trees (when $e$ was removed). To be more precise, denote by $w_e,w_e'$ the two vertices to which $e$ is incident. Edge $e$ partitions the vertices of $T$ into two subsets, which we call $Z_{e}^{1}$ and $Z_{e}^{2}$ defined by: 
\begin{eqnarray*}
Z_{e}^{1}\left(T\right)&:=&\{w\in S| e\text{ is contained in the unique }
(w_e,w)\text{-path in }T\}\\
Z_{e}^{2}\left(T\right)&:=&\{w\in S| e\text{ is contained in the unique }
(w_e',w)\text{-path in }T\}
\end{eqnarray*}
Example~\ref{ex:Ue1} visualizes this definition. 
We observe that edge $e$ occurs in all $|Z_{e}^{2}\left(T\right)|$ paths from any node $v \in Z_{e}^{1}\left(T\right)$ to any node $w \in Z_{e}^{2}\left(T\right)$. Note that the total number of paths in which $e$ occurs is $|Z_{e}^{1}\left(T\right)|\cdot|Z_{e}^{2}\left(T\right)|$. This fact is later used to compute $RC_S\left(T\right)$. 

\begin{algorithm}[!htbp]
\begin{algorithmic}[1]
\State $rc_S:=\{\infty,\dots,\infty\}$; //** is updated during the runtime of the algorithm.
\If{$u\in S$}
\State $z:=\{1,\dots,1\}$; //** is updated during the runtime of the algorithm.
\Else
\State $z:=\{0,\dots,0\}$;
\EndIf
\For{$t=1,\dots,|S|+2D_\omega'$}
\State within one time slot:
\newline \hspace*{0.8cm}\textbf{For each} $v\in L$ \textbf{such that} $t=|S|+2\cdot D_\omega'-\tau[v]$ \textbf{send} $\left(v,rc_S[v],z[v]\right)$ to
\newline \hspace*{0.8cm}$parent\_in\_T_{v}$;
\newline \hspace*{0.8cm}\textbf{receive} $\left(v_1,r_1,z_1\right)$ from neighbor $u_1$; //** $r_1$ equals $rc_S\left(T_{v_1},u_1\right)$,
\newline \hspace*{6.2cm}//** $z_1$ equals $Z_{\left(u,u_1\right)}^1\left(T_{v_1}\right)$
\newline \hspace*{0.8cm}\textbf{receive} $\left(v_2,r_2,z_2\right)$ from neighbor $u_2$; //** $r_2$ equals $rc_S\left(T_{v_2},u_2\right)$,
\newline \hspace*{6.2cm}//** $z_2$ equals $Z_{\left(u,u_2\right)}^1\left(T_{v_2}\right)$
\newline \hspace*{0.8cm}$\vdots$
\newline \hspace*{0.8cm}\textbf{receive} $\left(v_{\delta\left(u\right)},r_{\delta\left(u\right)},z_{\delta\left(u\right)}\right)$ from $u_{\delta\left(u\right)}$; //** $r_{\delta\left(u\right)}$ equals $rc_S\left(T_{v_{\delta\left(u\right)}},u_{\delta\left(u\right)}\right)$,
\newline \hspace*{6.3cm}//** $z_{\delta\left(u\right)}$ equals $Z_{\left(u,u_{\delta\left(u\right)}\right)}^1\left(T_{v_{\delta\left(u\right)}}\right)$
\For{$i=1,\dots,\delta\left(u\right)$}
\If{$v_i\neq\bot$}
\State $rc_S[v_i]:=rc_S[v_i]+r_i+2\omega_G\left(u,v\right)\cdot z_i\cdot \left(|S|-z_i\right)$;
\State $z[v]:=z[v]+z_i$;
\EndIf
\EndFor
\EndFor
\State //** Now $rc_S[u]$ equals $RC_S\left(T_u\right)$ in case that $u\in S$. Else it is $\infty$ and was never modified. 
\end{algorithmic}
\caption{Computing $RC_S(T_v)$ for each $v\in S$ alternative Part 2 (executed by node $u$)}\label{ALGRAND}
\vspace*{0.5cm}
\end{algorithm}

\begin{example}\label{ex:Ue1}
In Figure~\ref{fig:mrcst} we consider a graph $G$ and edge $e=(u_e,u_e')$ and assume $v$ and $u_e$ are elements of $S$. Then $Z_{e}^{2}\left(T\right)=\left\{u_e,v\right\}$ and  $Z_{e}^{1}\left(T\right)=S\backslash Z_{e}^{2}\left(T\right)$. Edge $e$ is part of all $|S|-2$ paths from $v$ to a node $u \in Z_{e}^{1}\left(T\right)$ and also on all $|S|-2$ paths from $u_e$ to all nodes $u\in Z_{e}^{1}\left(T\right)$. Thus in total $e$ occurs in $|Z_{e}^{1}\left(T\right)|\cdot|Z_{e}^{2}\left(T\right)|=\left(|S|-2\right)\cdot 2$ paths. 
\begin{figure}[ht]
\centering
\includegraphics{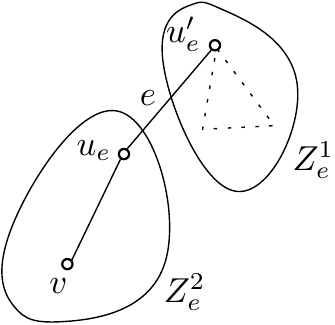}
\caption{Example of vertex sets $Z_{e}^{1}\left(T\right)$ and $Z_{e}^{2}\left(T\right)$ in a graph.}
\label{fig:mrcst}
\end{figure} 
\end{example}

\begin{lemma}\label{lem:lem}
For a tree $T$, the routing cost $RC_S(T)$ can be restated as $RC_S(T)=2\cdot\sum_{e \in T}|Z_{e}^{1}(T)|\cdot|Z_{e}^{2}(T)|\cdot  \omega_G(e)$.
\end{lemma}

\begin{proof}
We define a function $\delta$ indicating whether an edge is part of the unique path between two nodes in $T$. 
$$
\delta_{v,w}(e) := 
 \begin{cases}
1 : & $e$ \text{ is on the unique path from }
 $v$\text{ to }$w$\text{ in }$T$, \\
0: & \text{otherwise}.
 \end{cases}
$$
and restate
\begin{align*}
RC_S(T)& = \sum_{v,w \in S} \omega_T (v,w) = \sum_{v,w \in S} \sum_{e \in P_{v,w}} \omega_G(e) 
 = \sum_{v,w \in S} \sum_{e \in T} \delta_{v,w}(e) \cdot \omega_G(e) 
\\
& = \sum_{e \in T}\left(\omega_G(e)\cdot \sum_{v,w \in S} \delta_{v,w}(e) \right)
 =2\cdot\sum_{e \in T}\omega_G(e)\cdot|Z_{e}^{1}(T)|\cdot|Z_{e}^{2}(T)|
\end{align*}

Where we use in the last transformation the fact that $\sum_{v,w \in S} \delta_{v,w}(e)$ is the total number of $(v,w)$-paths which contain the edge $e$, which can be expressed as $ |Z_{e}^{1}(T)|\cdot|Z_{e}^{2}(T)|$ as noted in the text before Example~\ref{ex:Ue1}. 
\end{proof}

To formulate the definition of $RC_S(T)$ in this way helps us to argue that we can compute $RC_S(T)$ recursively in a bottom-up fashion for any $T$. To do so, we consider trees to be oriented such that we use the notion of child/parent. 

\begin{definition}[Subtree, partial routing cost]
Given a tree $T$, for each node $u$ in an oriented tree $T$, we define $T|_u$ to be the subtree of $T$ rooted in $u$ containing all descendants of $u$ in $T$. Denote by $V_v$ the vertices in $T|_v$. Given node $u$, denote by $rc_S\left(T,u\right)$ the part of the routing cost $RC_S\left(T\right)$ that is due to the edges in $T|_u$. We define $rc_S(T,u)$ in a recusive way. In case that $T|_u$ consists of only one node, $T|_u$ contains no edges that could contribute to $rc_S\left(T,u\right)$ and we set $rc_S\left(T,u\right):=0$. In case that $T|_u$ contains more than one node, we denote the children of $u$ in $T$ by $u_1,\dots,u_{\delta\left(u\right)-1}$ and define $rc_S\left(T,u\right):=\sum_{i=1}^{\delta\left(u\right)-1} rc_S\left(T,u_i\right) + 2\cdot\sum_{i=1}^{\delta\left(u\right)-1} \omega_G\left(u,u_i\right)\cdot |Z_{\left(u,u_i\right)}^1\left(T\right)|\cdot|Z_{\left(u,u_i\right)}^2\left(T\right)|$.
\end{definition}

Note that $rc_S(T,u)$ is a measure with respect to the routing cost in $T$ and thus different from $RC_S(T|_u)$. Besides $RC_S(T|_u)$ being undefined when $T|_u$ does not contain all nodes in $S$, $RC_S(T|_u)$ would take only routing cost within $T|_u$ into account.

We now formally prove that $rc_S\left(T,u\right)$ essentially describes the contribution of edges in subtree $T|_u$ to the total routing cost and conclude:
\begin{lemma}\label{lem:part-rc}
Let $T$ be a tree rooted in node $r$. Then $RC_S(T)=rc_S(T,r)$.
\end{lemma}
\begin{proof} We know due to Lemma~\ref{lem:lem} that 
\begin{align*}
rc_S(T,u)& = 2\cdot\sum_{e \in T|_u} \omega_G\left(e\right)\cdot|Z_e^1\left(T\right)|\cdot|Z_e^2\left(T\right)|.
\end{align*}
Observe that $T|_u$ consists of the subtrees $T|_{u_1},\dots,T|_{u_{\delta(u)-1}}$ induced by $u$'s children and the edges $\left(u,u_1\right),\dots,$ $\left(u,u_{\delta(u)-1}\right)$. Thus we can split the above term to be 
$$= \sum_{i=1}^{\delta\left(u\right)-1} 2\cdot\sum_{e \in T|_{u_i}} \omega_G\left(e\right)\cdot |Z_{e}^1\left(T\right)|\cdot|Z_{e}^2\left(T\right)| 
+ 2\cdot\sum_{i=1}^{\delta\left(u\right)-1} \omega_G\left(u,u_i\right)\cdot |Z_{\left(u,u_i\right)}^1\left(T\right)|\cdot|Z_{\left(u,u_i\right)}^2\left(T\right)|
$$
which in turn is
$$
\sum_{i=1}^{\delta\left(u\right)-1} rc_S\left(T,u_i\right) 
 + 2\cdot\sum_{i=1}^{\delta\left(u\right)-1} \omega_G\left(u,u_i\right)\cdot |Z_{\left(u,u_i\right)}^1\left(T\right)|\cdot|Z_{\left(u,u_i\right)}^2\left(T\right)|
$$
\end{proof}

Using this insight we are able to compute $RC_S\left(T_v\right)$ for all $v\in S$ in parallel recursively in a bottom-up fashion. This is by computing $rc_S\left(T_v,u\right)$ for each $u$ based on aggregating $rc_S\left(T_v,u_j\right)$ for each of $u$'s children. For each $v\in S$ these computations of $RC_S\left(T_v\right)$ run in parallel. A schedule on how to do these bottom-up computations in time $\BO\left(|S|+D_\omega\right)$ is provided by using the inverted entries of $\tau$. 

In more detail each node $u$ computes for each $v\in S$ the costs $rc_S(T_v,u)$ (stored in $rc_S[v]$) of its subtree of $T_v$ as well as the number of nodes in $T_v|_u$ (stored in $z[v]$ and sends this information to its parent in $T_v$. When we computed $T_v$ in Algorithm~\ref{alg:S-MRCT-Part-1}, we connected $u$ via edge $parent\_in\_T_v$ to $T_v$ at time $\tau[v]$. To avoid congestion we send information from $u$ to its parent in $T_v$ only at time $t=|S|+2D_\omega'-\tau[v]$ (Line 7). Note that this schedule differs from the one that is implied by the computation of the trees in the sense that now only edges in the tree are used, while more edges were scheduled while building the trees. The edges used now in time slot $t=|S|+2D_\omega'-\tau[v]$ are a subset of those scheduled at time $t=|S|+2D_\omega'-\tau[v]$ while constructing the trees, such that there is no congestion from this modification.

At the same time as $u$ sends, $u$ receives messages from its neighbors. E.g. neighbor $u_i$ might send $rc_S(T_{v'},{u_i})$ and $Z_{\left(u,u_i\right)}^1\left(T_{v'}\right)$ for another node $v'$. In Lines $8-11$ node $u$ updates its memory depending on the received values. In the end the node with ID $1$ computes $v:=arg\min_{v\in V}RC_S\left(T_v\right)$ via aggregation using $T_1$. Node $1$ informs the network that tree $T_v$ is a $2$-approximation to an $S$-MRCT.
\begin{theorem}\label{thm:exact_RC}
The algorithm presented in this section computes all $|S|$ values $RC_S(T_v)$ for each node $v\in S$ in time $\BO(|S|+D_\omega)$.
\end{theorem}
\begin{proof}
\textbf{Runtime:} The construction of the $|S|$ trees in Algorithm~\ref{alg:S-MRCT-Part-1} takes at most $\BO\left(|S|+D_\omega\right)$ rounds as stated in Lemma~\ref{lem:S-MRCT-Part-1-Runtime}. To forward/compute the costs from the leaves to the roots $v\in S$ in Algorithm~\ref{ALGRAND} takes $|S|+2D_\omega'$ since we just use the schedule $\tau$ of this length computed in Algorithm~\ref{alg:S-MRCT-Part-1}. Thus the total time used is $\BO\left(|S|+D_\omega\right)$.

\textbf{Correctness:} 
We consider time slot $|S|+2D_\omega'-\tau[v]$. If $u$ is a leaf of $T_v$, it sends $(v,0,1)$ to its parent in $T_v$ in case $u\in S$, else it sends $(v,0,0)$, which is correct. In case $u$ is not a leaf, each child $u_i$ has sent $rc_S\left(T_v,u_i\right)$ (stored in $r_i$) as well as $Z_{\left(u,u_i\right)}^1\left(T_{v}\right)$ (stored in $z_i$) to $u$ at an earlier point in time. This is true as time-stamp $\tau[v]$ stored in $u_i$ is always larger than time-stamp $\tau[v]$ stored in $u$, as $u_i$ is a child of $u$. Each time $u$ received some of these values from its children in $T_v$, it updated its memory according to Lemma~\ref{lem:part-rc} (Lines $8-11$ of Algorithm~\ref{ALGRAND}), leading to sending the correct values $rc_S\left(T_v,u\right)$ and $Z_{\left(parent\_in\_T_v,u\right)}^1(T_v)$ to its parent in $T_v$ at time $|S|+2D_\omega'-\tau[v]$. Thus in any case $u$ sends the correct values.

We conclude that each node $v\in S$ has computed $rc_S(T_v,v)=RC_S(T_v)$ after Algorithm~\ref{ALGRAND} has finished. 
\end{proof}

\section{Proofs of Main Results}\label{sec:final}
We put the tools of the previous sections together and prove the Theorems of Section~\ref{sec:intro}.
\begin{proof}(of Theorem~\ref{thm:mrct}).
First, Algorithms~\ref{alg:S-MRCT-Part-1} and~\ref{ALGRAND} are used to compute $RC_S(v)$ for each $v\in S$. For each such node $v$, the value $RC_S(v)$ is stored in node $v$ itself. A leader node (e.g. with lowest ID, which can be found in time $\BO(D_\omega)$) computes $u:=arg\min_{v\in V}RC_S(v)$ via aggregation using $T_l$, where $l$ is the leader node. As stated in Theorem~\ref{thm:det-ana} the tree $T_u$ is a $\left(2-2/|S|\right)$-approximation of a $S$-MRCT. The leader node informs the network that tree $T_u$ is a $\left(2-2/|S|\right)$-approximation to an $S$-MRCT.
The runtime follows from Lemma~\ref{lem:RC-comp} and the fact, that to determine $u$ by aggregating the corresponding minimum and to broadcast $u$ can be done in time $\BO(D_\omega)$.
\end{proof}
\begin{proof}(of Theorem~\ref{thm:mrct-rand}).
First we select a subset $S'\subseteq S$ of the size stated in Lemma~\ref{lem:rand-ana}. 
Each node joins a set $S''$ with probability $c\cdot s/n$, where $s$ is the (desired) size of $S'$ stated in Lemma~\ref{lem:rand-ana} and $c$ a constant depending on a Chernoff bound used now. Using such a Chernoff Bound, w.h.p. $S''$ is of size $c\cdot s$ or some constant $c\geq 1$. Now all IDs of nodes in $S''$ are sent to the leader who selects and broadcasts a subset $S'$ of the desired size among the IDs of $S''$.  
 
From now on the algorithm works exactly as in the proof of Theorem~\ref{thm:mrct}, except that the algorithm is run on $S'$ instead of $S$ (it computes and aggregates each $RC_S(v)$ for $v\in S'$ instead of $S$). As stated in Lemma~\ref{lem:rand-ana}, a tree $T_u$ is found that is a $\left(2-2/|S|+\beta(n,D)\right)$-approximation of an $S$-MRCT. The leader node informs the network that tree $T_u$ is a $\left(2-2/|S|+\beta(n,D)\right)$-approximation to an $S$-MRCT. Choosing $\beta(n,D):=\min\left\{\frac{\log n}{D},\alpha(n,D)\right\}$ yields the desired approximation ratio of $2-2/|S|+\min\left\{\frac{\log n}{D},\alpha(n,D)\right\}$, as stated in the Theorem.

\emph{Runtime analysis:} As $s=\left(\left\lceil\frac{2-2/|S|}{\beta(n,D)}\right\rceil+1\right)\cdot \ln n$, selecting a set $S''$ and deriving $S'$ can be done w.h.p. in time 
$$\BO(D+s)=\BO\left(D+\left(\left\lceil\frac{2-2/|S|}{\beta(n,D)}\right\rceil+1\right)\cdot \ln n\right)=\BO\left(D+\frac{\log n}{\beta(n,D)}\right),$$
which is $\BO\left(D+\frac{\log n}{\alpha(n,D)}\right)$ due to the choice of $\beta$. The same runtime follows from Lemma~\ref{lem:RC-comp} for computing the single source routing costs for all $v\in S'$. Combined with the fact that the aggregation and broadcast of $u$ can be done in time $\BO(D)$, the stated result is obtained.
\end{proof}

\section{Why $2$-Approximations Can't be Improved Cheap}\label{sec:cant}
The following example demonstrates a setting where an SP-tree yields a $2$-approximation to the routing cost of the underlying graph $G$, while no subgraph $H$ with $o(n^2)$ edges can yield a $(2-\varepsilon)$-approximation, which  demonstrates the strength of the tree that is able to provide a $2$-approximation while it has only $\BO(n)$ edges while . 
\begin{example} 
Let $G$ be the clique with uniform edge-weights $1$. For $S:=V$ we obtain that the routing cost $RC_S(G)$ is $n(n-1)$. Any SP-tree $T$ yields $RC_V(T)=(n-1)+2(n-1)(n-2)+(n-1)=2(n-1)^2$: the routing cost between the root $r$ and all other nodes is $(n-1)$. The routing cost of each of the remaining $n-1$ nodes $v\in V\setminus {r}$ to the nodes $u\in V\setminus {r,v}$ via paths of length $2$ is $2(n-1)(n-2)$. The routing costs from nodes $v\in V\setminus {r}$ to $r$ is $n-1$. Thus $RC_V(T)$ is a factor $2-2/|S|$ off from $RC_S(G)$. As all paths between two nodes using edges of $T$ are of length at most two, the only way to reduce the routing cost by a factor of $\varepsilon$ is to carefully add more than $\varepsilon (n-1)^2$ edges to the tree. Thus the total cost of such an approximation structure is a factor $\varepsilon (n-1)$ higher than the cost of a tree. 
\end{example}

\section*{Acknowledgment}
We would like to thank Benjamin Dissler and Mohsen Ghaffari for helpful discussions and insights.

\addcontentsline{toc}{section}{References} 
\bibliographystyle{plain}
\bibliography{references}

\end{document}